\documentclass[oldversion]{aa}
\usepackage{graphicx}
\usepackage{natbib}
\usepackage{hyperref}
\usepackage{longtable}
\usepackage{lscape}
\begin{document}

   \title{Binaries among low-mass stars in nearby young moving groups
\thanks{Based on observations collected at the European Southern Observatory, Chile (Programs 096.C-0243 and 097.C-0135).}
}


   \author{Markus Janson\inst{1} \and
          Stephen Durkan\inst{2} \and
          Stefan Hippler\inst{3} \and
          Xiaolin Dai\inst{3} \and
          Wolfgang Brandner\inst{3} \and
          Joshua Schlieder\inst{4} \and
          Micka{\"e}l Bonnefoy\inst{5} \and
          Thomas Henning\inst{3} 
	  }

   \offprints{Markus Janson}

   \institute{Department of Astronomy, Stockholm University, Stockholm, Sweden\\
              \email{markus.janson@astro.su.se}
         \and
             Astrophysics Research Center, Queens University Belfast, Belfast, Northern Ireland, UK\\
	\email{sdurkan01@qub.ac.uk}
	\and
	    Max Planck Institute for Astronomy, Heidelberg, Germany\\
	\email{hippler@mpia.de, xiaolin@mpia.de, brandner@mpia.de, henning@mpia.de}
	\and
	    NASA Exoplanet Science Institute, Caltech, Pasadena, California, USA\\
	\email{jschlied@ipac.caltech.edu}
	\and
	    Univ. Grenoble Alpes, IPAG, Grenoble, France\\
	\email{mickael.bonnefoy@univ-grenoble-alpes.fr}
             }

   \date{Received ---; accepted ---}

   \abstract{The solar galactic neighbourhood contains a number of young co-moving associations of stars (so-called `young moving groups') with ages of $\sim$10--150~Myr, which are prime targets for a range of scientific studies, including direct imaging planet searches. The late-type stellar population of such groups still remain in their pre-main sequence phase, and are thus well suited for purposes such as isochronal dating. Close binaries are particularly useful in this regard, since they allow for a model-independent dynamical mass determination. Here we present a dedicated effort to identify new close binaries in nearby young moving groups, through high-resolution imaging with the AstraLux Sur Lucky Imaging camera. We surveyed 181 targets, resulting in the detection of  61 companions or candidates, of which 38 are new discoveries. An interesting example of such a case is 2MASS~J00302572-6236015~AB, which is a high-probability member of the Tucana-Horologium moving group, and has an estimated orbital period of less than 10 years. Among the previously known objects is a serendipitous detection of the deuterium burning boundary circumbinary companion 2MASS~J01033563-5515561~(AB)b in the $z^{\prime}$-band, thereby extending the spectral coverage for this object down to near-visible wavelengths. }


\titlerunning{Young low-mass binaries}
\authorrunning{M. Janson et al.}

   \maketitle
%

\section{Introduction}
\label{s:intro}

Young Moving Groups (YMGs) are associations of stars that, in addition to having individual indications of youth, are clustered in phase space and therefore generally assumed to originate from a common birth cluster \citep[e.g.][]{torres2000,zuckerman2001}. Thus, they can be expected to be approximately co-eval, which opens up for a range of scientific opportunities that are otherwise unattainable. For instance, statistical age estimators can be applied to a large number of stars in a YMG in order to average out the scatter and improve the precision of the age, and conversely, if the age of individual stars can be determined with particularly good accuracy, this can feed back to the age estimation of all other stars that are associated with the same YMG. The age is a fundamental parameter for many purposes in stellar science, but one of the main reasons for why it has received particular attention in recent years is its relevance for exoplanet research. Since planets are maximally hot directly after formation and subsequently cool gradually for the rest of their lifetime (or until they reach a thermal equilibrium due to illumination from a parent star), the interpretation of a directly imaged planet depends crucially on understanding its age \citep[e.g.][]{marois2008,lagrange2010,kuzuhara2013}. Since YMGs are additionally prime targets for direct imaging surveys due to their youth and proximity \citep[e.g.][]{chauvin2010,biller2013,brandt2014}, understanding the ages of YMGs is intimately coupled to the understanding of directly imaged planets.

M-type stars have rather long pre-main sequence lifetimes of $\sim$100 Myr \citep[e.g.][]{baraffe1998}. As a result, when such stars reside in YMGs, they are typically still evolving through the pre-main sequence phase, and thus they can potentially be isochronally dated to a good level of precision. Such isochronal dating can in principle be performed using a $T_{\rm eff}$ versus $L_{\rm bol}$ relationship \citep[e.g.][]{janson2007}, but the temperature scale of M-dwarfs is highly uncertain and thus a potential cause of systematic error. A more robust analysis can be based on using an $M_{\rm star}$ versus  $L_{\rm bol}$ relationship instead. This has been achieved in a few cases \citep[e.g.][]{bonnefoy2009,kohler2013,montet2015a}, but more examples would be highly beneficial for covering more YMGs and more reference cases in each YMG to test for robustness, coevality within YMGs, etc. In previous high-resolution imaging surveys, we have identified a large number of binaries in young systems \citep{bergfors2010,janson2012a}. However, not all of these can be associated to known YMGs, and only a subset of the discovered binaries have orbital periods that are short enough so that robust orbital parameters can be estimated in a reasonably rapid timeframe \citep{janson2014a}. In order to increase the yield of high-utility binaries, it would be most efficient to target stars that have already been identified as YMG members.

Here, we present the results of a high-resolution imaging survey of low-mass stars that have been identified as probable members of nearby YMGs \citep[][]{malo2013,malo2014,kraus2014}, primarily motivated by the reasoning above. In the following, we will first present the selected sample of stars in Sect. \ref{s:targets} and the acquisition and reduction of the imaging data in Sect. \ref{s:obs}. We will then discuss the results for the sample at large in Sect. \ref{s:multiplicity}, pay special attention to the `planetary mass' companion 2MASS J01033563-5515561(AB)b in Sect. \ref{s:2m0103} and note peculiarities of other targets in Sect. \ref{s:notes}. Finally, we will summarize the results of the survey in Sect. \ref{s:summary}.

\section{Target sample}
\label{s:targets}

Our targets were selected from three catalogues of late-type stars identified as high-probability members of nearby YMGs \citep[][]{malo2013,malo2014,kraus2014} that had not been previously monitored with AstraLux. In principle, six observational parameters are required to exactly relate the $XYZUVW$ of the target to that of various moving groups it could conceivably be associated to, but since late-type stars are faint at visible-light wavelengths, most potential YMG members lack a parallactic distance, and many lack a radial velocity measurement, so a probabilistic estimation must sometimes be made on the basis of only four parameters (right ascension, declination, and proper motion along both these directions). Nonetheless, the Bayesian estimations made in e.g. \citet{malo2014} using the so-called BANYAN code can often distinguish field objects from moving group members with such a constrained parameter set with quite high probabilities, provided that the priors are trustworthy. 

For the purpose of scheduling and executing the observations, we took the membership assignments in the aforementioned catalogues at face value. However, since then, the first \textit{Gaia} data release has been presented \citep{brown2016}, including the Tycho-Gaia Astrometric Solution \citep[TGAS, see][]{michalik2015}. This contains new parallaxes and improved proper motions for 29 of the observed targets. Hence, for these cases we re-evaluated YMG membership using the BANYAN II tool \citep{gagne2014} with the updated astrometry from \textit{Gaia}. In many cases, the \textit{Gaia} parallactic distance was remarkably close to the kinematic distance prediction yielded by BANYAN when no measured parallax is provided. Some particularly notable cases are mentioned in Sect. \ref{s:notes}. In this way, we found that the YMG membership hypothesis is supported (and strengthened) with the addition of a \textit{Gaia} parallax in 17 of the 29 cases. However, in the other 12 cases, membership could not be supported, with the YMG probability going down to effectively 0\% in some cases. This relatively high rejection rate should be taken as a caution that the other targets, for which no parallactic distance exists yet, should still be taken only as candidate YMG members, rather than bona fide members. Still, the fact that a majority of the re-assessed cases were verified as YMG members can be seen as an indication that most of the other candidate assignments are probably also correct. In due time, \textit{Gaia} will provide accurate distances to all of these targets, such that membership can be more rubustly assessed across the board.

The sample properties are listed in Table \ref{t:sample}. 


\section{Observations and Data Reduction}
\label{s:obs}

The observations of this survey were taken during three separate runs during the European Southern Observatories (ESO) observing periods P96 (program 096.C-0243) and P97 (program 097.C-0135). Two of the runs were scheduled in P96, with four nights spanning 24--27 Nov 2015 and another four nights spanning 24--27 Dec 2015. Unfortunately, the Nov run suffered from very poor conditions with clouds, strong winds, and poor seeing, so only a few targets could be usefully observed for the survey on the night of Nov 27. The Dec run was considerably better and yielded useful data every night. In P97, the single four-night run was scheduled for 16--19 May 2016. One of these nights, May 17, was very productive and offered a clear sky and good seeing for most of the night, but like in the Nov run, the rest of the time was clouded out. 

For all of the observations, we used the AstraLux Sur Lucky Imaging camera \citep{hippler2009} mounted at the guest instrument port of the 3.5m New Technology Telescope (NTT) in La Silla. The $z^{\prime}$ filter was used, and typically 10000 frames with individual integration times of a few tens of ms were acquired for each target. In total, we observed 181 separate targets during the available clear nights. Nine of these were observed twice, since some of the binary candidates detected in the late 2015 runs were re-observed in the May 2016 run. Generally, a subarray readout of $256 \times 256$ pixels was used since we are primarily interested in close binaries, but in the case of wide binaries, where both components may be interesting or where it was not immediately obvious to the observer which star was the actual target, the full $512 \times 512$ detector was read out in order to maintain both components firmly within the field of view. The field of view for the full frame is approximately $16^{\prime \prime} \times 16^{\prime \prime}$.

Data reduction was done in an identical way to our previous AstraLux surveys \citep[e.g.][]{janson2012a,janson2014a}. The reduction pipeline \citep{hormuth2008} produces several Lucky Imaging outputs with different levels of selection (e.g. all frames used, the 10\% best seeing frames used, the 1\% best seeing frames used etc). For our purposes, we consistently chose a 10\% selection for further analysis as it provides a good trade-off between resolution and sensitivity. The final pixel scale as well as the True North orientation were determined individually for the different runs by observing clusters that have been previously observed with the Hubble Space Telescope (HST) for astrometric reference. For the Nov 2015 and Dec 2015 runs, we used Trapezium for this purpose with HST coordinates from \citet{mccaughrean1994}, and for the May 2016 run, we used M15 with HST coordinates from \citet{marel2002}. In this way, we found pixel scales of 15.19 mas/pixel for the Nov 2015 run, 15.20 mas/pixel for the Dec 2015 run, and 15.27 mas/pixel for the May 2016 run. Likewise, we derived True North orientations of 2.17 deg for Nov 2015, 2.40 deg for Dec 2015, and 3.04 deg for May 2016. The calibration uncertainties range 0.06--0.13 mas/pixel in pixel scale and 0.16--0.30 deg in True North orientation.

From the output images, relative astrometry and photometry was derived for any pair of stars that could be identified in the images. This was again done as in our previous studies, with aperture photometry and Gaussian centroiding in the case of wide separation pairs, and Point Spread Function (PSF) fitting for close separation pairs, where the PSFs of single stars were used as references. Triple systems consisting of a close pair and a third wide component need special attention in this regard. We handled these by choosing an aperture large enough to encompass both components of the close pair, thus acquiring an AB--C (or A--BC) relative photometry. The relative photometry of the close pair was then determined through PSF fitting as usual, and from this the relative photometry between the brighter component of the pair and the wider component could be derived. Near-equal brightness pairs are subject to the so-called `false triple' effect \citep{law2006thesis}, in which the pair of components map onto three apparent PSFs in the final image, also require special attention. In the context of astrometry, there is a 180 deg ambiguity in the position angle of such cases, since it cannot be uniquely determined which is which of the two binary components (this is an issue for nearly equal-brightness components even in the absence of a false triple effect). In the context of photometry, we have shown in previous papers \citep{janson2012a,janson2014b} that the photometric determinations on false triple pairs are biased, hence for this study, we simply omit a relative photometric analysis of such systems.

\section{Results and discussion}
\label{s:results}

\subsection{Multiplicity in the sample}
\label{s:multiplicity}

\begin{figure}[htb]
\centering
\includegraphics[width=8cm]{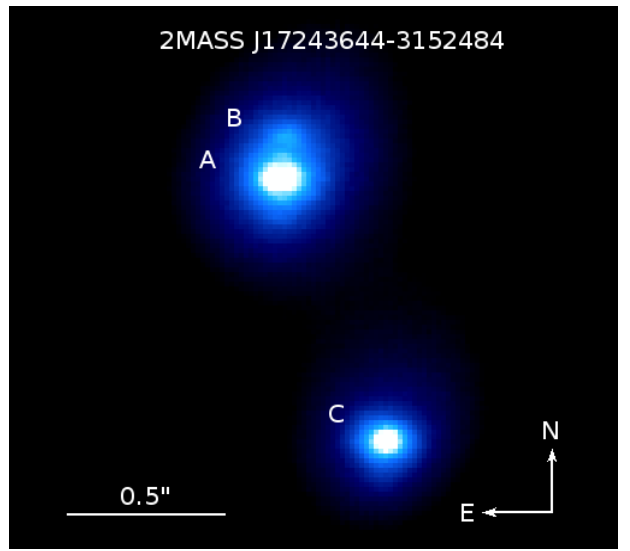}
\caption{Image of the J17243644-3152484 triple system, resolved by AstraLux for the first time. The YMG classification of this system is unclear, as discussed in Sect. \ref{s:notes}.}
\label{f:im1724}
\end{figure}

Among our sample of 181 targets, we detect 61 candidate companions, of which 23 have been previously reported in the Washington Double Star catalogue \citep{mason2001} and other sources cited individually in this article. The other 38 are, to our knowledge, new detections. An example of the new detections, the triple system J17243644-3152484, is shown in Fig. \ref{f:im1724}. 

The astrometric properties of the multiple systems are summarized in Table \ref{t:astro}. Of the 23 companions that have been previously observed, each one could be confirmed to share a common proper motion (CPM) with the primary. This is consistent with our experience from previous surveys \citep[e.g.][]{janson2012a}, that the contamination frequency is low for typical targets (i.e., targets that are not very close to the galactic plane). While we do have second epoch observations for nine companions of which eight are new, the proper motions for several of them are too small to be statistically significantly measurable over a $\sim$5~month baseline. J12092998-7505400~AB could nonetheless be confirmed as CPM with $>$10$\sigma$ confidence. While not yet formally noted as CPM in the table, J07343426-2401353~AB and J07523324-6436308~BC could be seen as marginally confirmed CPM at 3--4$\sigma$. For the targets that neither have literature epochs nor could be confirmed as CPM, we made an additional test by checking digital sky surveys for historical epochs for the targets. This is similar to the checks performed in \citet{janson2014b}, as well as in other recent surveys \citep[e.g.][]{montet2015b,schlieder2016}. The best source for such checks is typically the Palomar Observatory Sky Survey (POSS), which includes images covering a large fraction of the sky from epochs around 1950, thus providing a sufficient baseline for targets with proper motions of $\sim$100~mas/yr to move by several arcseconds on the sky, which is usually more than the seeing-limited resolution of the images. Thus, background contaminants in the AstraLux image may be separated from the primary in such archival images for sufficiently fast-moving targets. Unfortunately, most of our targets are too far south to be included in POSS. In these cases, the best option is typically the so-called SERC survey from the UK Schmidt telescope, which has first epoch data from typically the late 1970s or early 1980s. Furthermore, some of our targets with candidate companions have a quite slow proper motion, down to 11 mas/yr in the slowest case. In order to be able to draw confident conclusions, we set the threshold that 3$^{\prime \prime}$ of motion must have occurred between the archival epoch and the AstraLux data for the archival epoch to be useful. This leaves 15 targets that can be usefully checked. Of these, there are 10 cases in which no candidate background star can be seen in the archival image. This strongly implies that the respective candidates are physical companions. In this context it should still be considered that the filters of the two epochs are quite different; the archival epochs are at shorter wavelengths than the AstraLux data. Hence, a hypothetical very red background source may escape detection in the archival data, in which case it would be erroneous to label the candidate as CPM. We consider that a positive detection is needed for a final confirmation of CPM, and thus we label these candidates as `implied CPM' rather than fully confirmed CPM. This is marked with the label `I' in Table \ref{t:astro}. In the remaining five cases, the separation between the primary and candidate was large enough that that they could be recognized as partially resolved in the archival images, such that a real CPM test could be performed. In this way, we could confirm three additional CPM companions (J00302572-6236015~C, J18450097-1409053~B, and J23204705-6723209~B), and two background contaminants (J00514081-5913320~B and J23332198-1240072~B). The relatively large rejection fraction is due to the fact that these are the widest among the candidates, which is where the contaminants are expected to reside. Of the candidates that remain unverified, we expect that three (J05111098-4903597~B and J16572029-5343316~B and C) are possible/probable background contaminants, but the rest are very close and/or very bright relative to the primary, hence they are all probable companions. Obviously, common proper motion tests over a longer timescale will be required to verify this beyond reasonable doubt in the individual cases.

Photometry and derived quantities are shown in Table \ref{t:phot}. The relative photometry was used to estimate individual masses using theoretical mass-luminosity relationships for young stars. Each target system was assigned an age based on their YMG membership assignment as shown in Table \ref{t:sample}. The ages for the AB Dor (ABMG, $\sim$150 Myr), $\beta$~Pic (bPMG, $\sim$25~Myr), Carina (CAR, $\sim$45~Myr), Columba (COL, $\sim$40~Myr), Tucana-Horologium (THA, $\sim$45~Myr) and TW Hya (TWA, $\sim$10~Myr) YMGs were adopted from \citet{bell2015}. Argus is considered a questionable association by \citet{bell2015}, but is used in the BANYAN framework, so we assign it an age of 40~Myr following \citet{torres2008}. For a given star with a given age, we adopt the evolutionary models of \citet{baraffe2015} to make predictions of the masses from the measured brightnesses. The combination of masses for the two components of a binary pair that minimizes the RMS residuals of the $\Delta z^{\prime}$ from our data and the total $J$-band magnitude from 2MASS \citep{skrutskie2006} is adopted for this purpose. The total mass $m_{\rm tot}$ of the binary in combination with an estimated semi-major axis $a$ can then in turn be used to make a tentative prediction for the orbital period $P$ of the pair, since $P \sim m_{\rm tot}^{-1/2} a^{3/2}$. An estimation for the semi-major axis can be acquired simply by considering that for sensible eccentricity distributions, the ratio between average projected separation and semi-major axis of a binary population is close to unity \citep{brandeker2006,bonavita2016}. In other words, the expectation value for the semi-major axis of a binary with no additional information about its orbit is approximately equal to its instantaneous projected separation. The period estimations resulting from this analysis are shown in Table \ref{t:phot}. These estimations are subject to large uncertainties, for a range of reasons such as systematic uncertainties in the models and ages, as well as the broad scatter in the translation between projected separation and semi-major axis. However, prior to measuring the actual orbits, this is the best that can be done with the data at hand, so it is a useful procedure for determining which systems are the most promising to follow up with a relatively high cadence for an orbital determination within a realistic timescale. Indeed, identifying such systems is the primary purpose of this study.

As mentioned previously, the false triple binaries are unsuitable for relative photometry and thus not included in Table \ref{t:phot}, but it is still of interest to acquire approximate period estimations. For this purpose, we apply a simplified procedure of assuming that the components are approximately equal brightness, and estimate the mass based on total $J$-band magnitude alone. Through this procedure, we acquire estimations of 8~yr for J00302572-6236015~AB, 43~yr for J01033563-5515561~AB, 9~yr for J02303239-4342232~AB, 295~yr for J04475779-5035200~AB, 111~yr for J12092998-7505400~AB, 89~yr for J17130733-8552105~AB, 39~yr for J20223306-2927499~AB, and 91~yr for J21342935-1840372~AB. In total among the pairs studied in this survey, this means that 9 pairs have an estimated orbital period $<$40~years, which we deem as the relevant cut-off timescale for sufficiently rapid orbits to motivate regular astrometric monitoring. We particularly note the two pairs J00302572-6236015~AB and J02303239-4342232~AB, which both have very rapid estimated orbital timescales of $<$10~years. J00302572-6236015 is identified as a THA member in \citet{kraus2014}, and a BANYAN II check supports this at a very high level of probability (99.7\% THA member with standard priors), though a parallactic distance is still required for it to formally be regarded as a bona fide member. J02303239-4342232 has been extensively studied in a YMG context \citep[e.g.][]{torres2008,schlieder2010}; in this study our baseline YMG assumption was based on the COL classification in \citet{malo2014}. Since then, a parallax has become available in TGAS which we could use to verify this membership in a BANYAN II analysis. We find that with standard priors, the probability of COL membership is 82.6\%, with a small alternative probability of 15.2\% that it is a member of THA instead. The probability of it being a field contaminant is only 1.9\%. In other words, it is very likely to be a genuine YMG member. We therefore consider J00302572-6236015~AB and J02303239-4342232~AB to be the very top priorities for astrometric follow-up out of the 61 examined in this work.

\subsection{The case of 2M0103}
\label{s:2m0103}

\begin{figure}[htb]
\centering
\includegraphics[width=8cm]{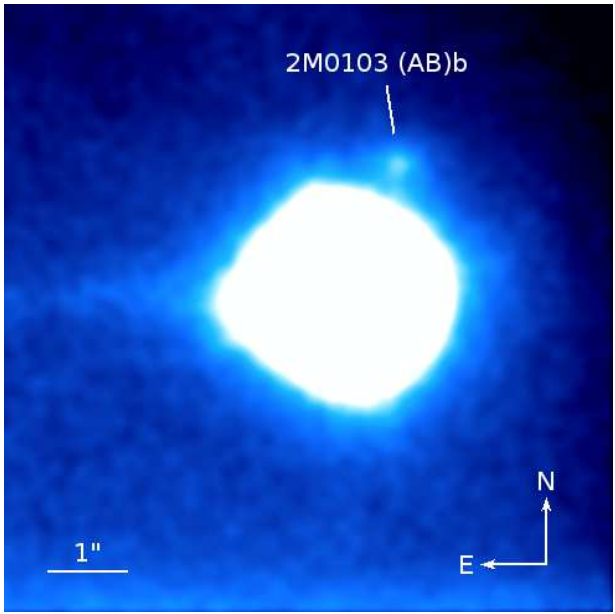}
\caption{AstraLux $z^{\prime}$ image of the 2M0103 system. Smoothing with a Gaussian kernel has been applied to better show the faint very low-mass companion first reported in \citet{delorme2013}.}
\label{f:im0103}
\end{figure}

An interesting ancillary outcome of our survey was the re-detection of J01033563-5515561(AB)b (system identifier hereafter abbreviated as 2M0103) originally reported in \citet{delorme2013}. Our image, which is displayed with a high stretch and has been smoothed with a Gaussian kernel of 20 pixel (304 mas) FWHM, can be seen in Fig. \ref{f:im0103}. 2M0103(AB)b was estimated in \citet{delorme2013} to be a $\sim$12--14~$M_{\rm jup}$ object at a relatively wide orbit ($\sim$84~AU projected separation) around a close low-mass stellar binary pair. It can therefore in some sense be regarded as a possible circumbinary planet, although it seems unlikely that it would have formed through standard planet formation mechanisms, at least at its present location. The targets of our survey were selected solely on the basis of probable late-type YMG members that had not previously been observed by AstraLux, hence the inclusion of this target in the survey was entirely coincidental. Only after we had discovered the point source in the AstraLux images, and a detailed literature search was made for 2M0103, was the history of the system realized.

Special considerations are required to constrain the properties of 2M0103(AB)b, because the conditions were non-optimal during the observation and the companion is faint and embedded in the PSF halo of the central binary, and also because the central binary exhibits a false triple effect, further complicating matters. We estimate relative photometry by using 15 pixel circular apertures around the primary pair and around the faint companion. The contribution of the PSF halo of the primary pair was estimated by taking the mean of the aperture flux at four different locations at the same separation from the pair as the companion. The points chosen for this purpose were due North, South, and West from the primary pair, as well as the point directly opposite to the companion's location. The Eastward direction was omitted for this purpose, since the PSF halo has a coma-like extension in that direction. In this way, we found that the AB--b contrast is $\Delta z^{\prime} = 6.5 \pm 0.3$~mag. This is the only photometric quantity that it directly measurable in the data due to the false triple effect of the primary, but if we assume that the A and B components have approximately equal brightness, which holds true at longer wavelengths \citep{delorme2013}, then the A--b contrast can be expected to be $\sim$5.7~mag.

\begin{figure*}[htb]
\centering
\includegraphics[width=18cm]{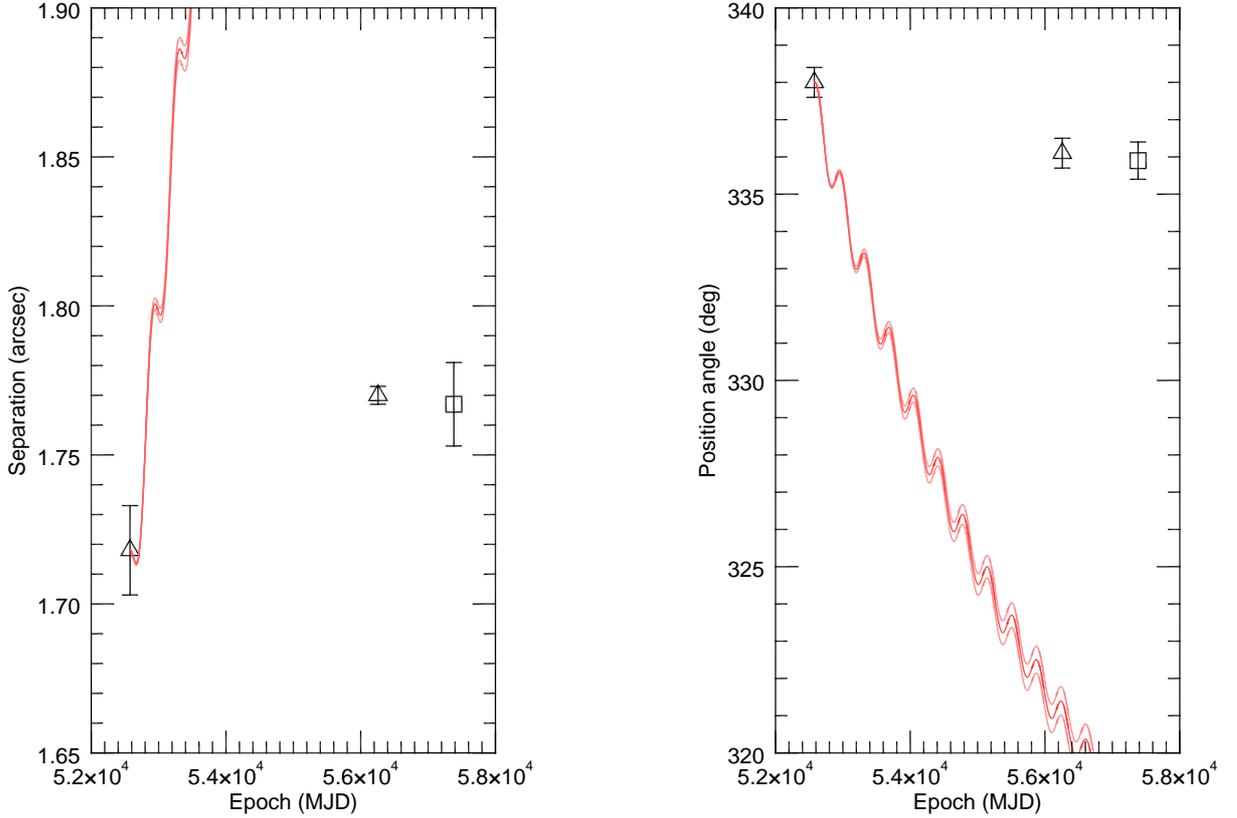}
\caption{Separation (left) and position angle (right) of 2M0103(AB)b relative to the AB barycenter as function of time. Triangles are data points from \citet{delorme2013} and the square in each panel is the astrometry from our AstraLux image. Red lines correspond to the motion that 2M0103(AB)b would exhibit if it were a static background star, which can be firmly excluded. There is also significant orbital motion between the first epoch in 2002 and the later epochs.}
\label{f:cpm0103}
\end{figure*}

For relative astrometry, we simply estimate the locations of the companion and the primary pair barycenter by eye, adopting a $\pm$1 pixel uncertainty. This gives a separation of $1.767 \pm 0.014$ arcsec and a position angle of $335.9 \pm 0.5$ deg. As can be seen in Fig. \ref{f:cpm0103}, the position in our Dec 2015 image is consistent with that in the Nov 2012 image of \citet{delorme2013}, and also confirms that orbital motion has occurred since their Oct 2002 epoch. Given the long baseline and relatively high proper motion of 2M0103, CPM is thoroughly obvious.

While a detailed analysis of the physical properties of 2M0103(AB)b is beyond the scope of this paper, it is interesting to assess whether the observational properties that we derive are consistent with the conclusions about the object properties in \citet{delorme2013}. In particular, we may ask whether our photometry is consistent with the $\sim$12--14~$M_{\rm jup}$ mass derived there. To test this, we first take the system age of 30~Myr adopted by \citet{delorme2013} at face value, and use the BT-SETTL models \citep{baraffe2015} for an isochronal mass-brightness conversion. Given a primary mass of $\sim$0.2~$M_{\rm sun}$, the models predict $M_{\rm z, A} = 8.3$~mag, and with an A--b contrast of 5.7~mag, this gives $M_{\rm z, b} = 14.0$~mag, which in turn gives a mass prediction of 13--14~$M_{\rm jup}$ from linear interpolation of public BT-SETTL grids. This is fully consistent with the previous value derived for the system. Hence, the data give a consistent picture of a companion spectral energy distribution (SED) for a $\sim$2000~K object, which at an age of 30~Myr corresponds to a mass in the range of the deuterium burning limit. However, a special note needs to be made about the age. The age assignment is based on THA membership in \citet{delorme2013}, so as a first check it is useful to test if we can support this conclusion. Our input sample classification for 2M0103 is THA, based on \citet{kraus2014}. We have re-assessed this membership using the BANYAN II tool, which supports this classification at a very high probability of 99.95\%, so the membership indeed seems to be very robust. A remaining issue, then, is what age THA actually has. The 30~Myr adopted in \citet{delorme2013} comes from \citet{torres2008}, but in this paper we have used the \citet{bell2015} ages, which for THA gives a somewhat older age range of $\sim$40--50~Myr. This corresponds to a BT-SETTL-predicted mass range of $\sim$15--20~$M_{\rm jup}$. Thus, we conclude that our data imply an SED consistent with the longer-wavelength data of \citet{delorme2013}, and that our analysis supports the classification of the system as a THA member, but we note that uncertainties in the specific age of THA broadens our error bars in mass relative to the discovery paper.

It is rather rare for such a planet-like object to be detected at such a short wavelength range as in $z^{\prime}$-band. Fomalhaut~b has been detected at yet shorter wavelengths \citep{kalas2008}, but its observed flux does not arise from a photosphere \citep[e.g.][]{janson2012b}. H$\alpha$ emission related to accreting planets may have been observed in the LkCa~15 system \citep{sallum2015}, but it remains unclear if the planets themselves have been imaged \citep{kraus2012,thalmann2016}. It is all the more unusual considering the rather modest size of the 3.5m NTT. While the use of Lucky Imaging with AstraLux was very useful in recovering this object, its detection is arguably less a question of instrumental capabilities than a question of target properties. It is the rare combination of proximity and youth that causes such a low-mass object as 2M0103(AB)b to be hot and luminous enough that it emits a non-negligible amount of radiation shortward of 1~$\mu$m. This underlines that studies of young moving groups with even modestly resource demanding instrument such as AstraLux have the potential for yielding very low-mass discoveries. In this context, it also emphasizes the importance of following up the candidate companions yielded by the survey over longer timescales to test for CPM. While the faintest candidates in our survey are by far the most likely to be background contaminants, they are also potentially the most interesting individual objects provided by the survey.

\subsection{Individual target notes}
\label{s:notes}

In this section, we make brief notes about a few targets that deserve or require particular attention.

\textbf{J02303239-4342232: }
This system was already discussed in Sect. \ref{s:multiplicity} due to the short estimated orbit of the AB pair. Here we additionally point out that there is a wide common proper motion companion at a separation of 173$^{\prime \prime}$ \citep{alonso2015}. Hence, the system is at least triple.

\textbf{J02442137+1057411: }
The YMG membership of this star was classified as ambiguous in \citet{malo2013}, but since we observe it as being multiple, it is preferable with a more specific designation if that can be determined, for the age input to the mass estimation of the binary components. Hence, we used the BANYAN II tool to see if some clearer YMG preference could be derived. In this analysis, J02442137+1057411 was assigned a 93.4\% probability of being a member of the $\beta$~Pic moving group with the standard prior. We thus consider it as a bPMG member here.

\textbf{J05015881+0958587: }
In addition to the wide component seen in the AstraLux images, which was originally reported in \citet{henry1997} and confirmed in \citet{delfosse1999}, the primary in the system also has a spectroscopic binary companion \citep{delfosse1999}. With a period of only 12 days, the spectroscopic pair is however significantly too close to be resolved in our images.

\textbf{J05064991-2135091: }
Two of the three components of this system have individual 2MASS identifiers: J05064991-2135091 and J05064946-2135038. The latter has a close companion which is the third component of the system. We used the full AstraLux frame instead of subarray readout for this target, such that both wide components could be included.

\textbf{J06153953-8433115: }
YMG membership of J06153953-8433115 is labelled as `ambiguous' in \citet{malo2013}, but since we resolve it as a binary, it is desirable to attain a more detailed view of YMG membership for age and mass assignment purposes. This turns out to be a complicated issue for this target -- BANYAN II gives only a 38.8\% probability of bPMG membership versus 59.9\% for the field with the standard priors. However, if we chose the priors to constrain the age to be $<$1~Gyr, the probability for bPMG membership goes up to 65.9\%. The latter choice is arguably appropriate, since the star is part of the \citet{riaz2006} sample which is indeed expected to be $<$1~Gyr. Hence, we adopt the bPMG age for this target, although we note that the evidence for such a membership is weak relative to most other targets in the survey.

\textbf{J11211723-3446454: }
The two components of this system have individual 2MASS identifiers, J11211723-3446454 and J11211745-3446497. The full-frame field of view of AstraLux Sur was used to fit both components in simultaneously.

\textbf{J14142141-1521215: }
In addition to the three components visible in the AstraLux field of view, there is a wide common proper motion companion at 65$^{\prime \prime}$ reported in e.g. \citet{alonso2015}, with identifier 2MASS~J14141700-1521125. This wide object happens to be part of the CASTOFFS survey \citep[][and Schlieder et al., in prep.]{schlieder2012,schlieder2015} and thus has FEROS spectroscopy dedicated to it. The spectrum shows signs of high activity, with nearly all of the Balmer series in emission, and very broad lines ($v \sin i = 41 \pm 4$~km/s). This lends further support to the kinematic indication that the system is young. Exactly how young remains a factor of uncertainty -- the classification of J14142141-1521215 in \citet{malo2013} is for the $\beta$~Pic moving group, but J14141700-1521125 fits better to the AB Dor YMG.

\textbf{J15244849-4929473: }
J15244849-4929473 is single in the AstraLux field of view, but is noted as a single-line spectroscopic binary in \citet{malo2014}. Since no constraints are given on the orbital period, it cannot be assessed whether the unseen companion might become visible to AstraLux during some phase of its orbit.

\textbf{J17165072-3007104: }
While this star was identified as a possible ambiguous YMG member in \citep{malo2013}, our re-analysis with BANYAN II implies that this is a young field star with essentially no chance of being part of any of the identified YMGs, for any sensible choice of priors. For mass estimations of this resolved binary, we thus arbitrarily chose an age of 150~Myr, which is the upper end of the YMG ages used in this study.

\textbf{J17243644-3152484: }
This is essentially an identical case as for J17165072-3007104 discussed above: Our BANYAN II re-analysis implies that this star is consistent with a young field star, but not possible to associate directly with any of the known young moving groups. As for J17165072-3007104, we use a tentative age of 150~Myr.

\textbf{J18141047-3247344: }
Also known as V4046 Sgr, J18141047-3247344 is a double-lined spectroscopic binary with a period of 2.4 days \citep{quast2000}. Furthermore, several studies \citep{kastner2011,alonso2015} have noted that it shares a common proper motion with J18142207-3246100, which is individually observed as a separate target in our survey. The separation between the two components is 169$^{\prime \prime}$. Neither of these are however in the detectability range of AstraLux, in which no further companions are identified.

\textbf{J19560294-3207186: }
The two stars J19560438-3207376 and J19560294-3207186, which were separately observed with AstraLux, form a wide binary pair with a separation of 26$^{\prime \prime}$. As already noted in \citet{bowler2015}, J19560294-3207186 is itself a close pair, which we confirm in this survey. Furthermore, J19560294-3207186 is noted as a double-lined spectroscopic binary in \citet{elliott2016}, implying that there could be a third component in this close sub-system, and thus a component of a quadruple system in total. The system is part of the CASTOFFS survey, in which the spectroscopy in fact shows three distinct lines in both absorption as well as in emission (H$\alpha$ and Ca II H) for J19560294-3207186, supporting the quadruple nature of the system.

\section{Summary and Conclusions}
\label{s:summary}

In this paper, we have presented high-resolution imaging observations of 181 late-type candidate members of nearby YMGs, with the purpose of identifying new binaries that can potentially be used for a range of calibration and age determination purposes. We discovered 61 candidate companions, of which 23 were previously known and the other 38 are new detections. Of the previously known companions, the most notable object is 2M0103(AB)b, which was reported as a $\sim$12--14~$M_{\rm jup}$ wide companion to a close pair of M-dwarfs \citep{delorme2013}. Our analysis of the $z^{\prime}$ photometry acquired in this work supports the general conclusions in \citet{delorme2013}, but we note that the upper mass limit needs to be extended up to 20~$M_{\rm jup}$ when accounting for the full possible age range of the Tuc-Hor association quoted in the recent literature \citep{torres2008,bell2015}. Of the companions in total, 9 have estimated orbital periods less than 40 years, and are thus important targets for astrometric follow-ups to constrain their orbital motions. 

We also used \textit{Gaia} data from the recent TGAS release to re-analyze YMG membership for those targets that have new parallactic distances which weren't available in the original classifications of the targets. We find that in the majority of cases (17 of 29), the YMG could be confirmed with the new information. Still, in a significant fraction of cases (12 of 29), YMG membership was rejected with the new information. Hence, any individual YMG candidate member without a parallactic distance should be treated with a certain degree of caution. Over the next few years, \textit{Gaia} will provide distances for all of the targets studied here, thus firmly identifying which systems are bona fide members and which are not. It is also likely to yield more YMG members that had previously been missed, and perhaps new nearby YMGs altogether. Finally, while the closest binaries resolved here will be too close to resolve with the much smaller \textit{Gaia}, absolute astrometry of the unresolved system can still yield valuable constraints of the orbits when combined with resolved imaging from the ground.

\begin{acknowledgements}
M.J. gratefully acknowledges funding from the Knut and Alice Wallenberg Foundation. S.D. acknowledges support from the Northern Ireland Department of Education and Learning. The authors thank the ESO staff for their efficient support, S. Ciceri for transatlantic data transportation, and the anonymous referee for useful suggestions. This study made use of the CDS services SIMBAD and VizieR, the SAO/NASA ADS service, data from the ESA mission \textit{Gaia}, and digitized archival data from the Anglo-Australian Observatory.
\end{acknowledgements}

\onecolumn


{\scriptsize
\begin{longtable}{llllrrrrlr}
\caption{Summary of the target sample.}\\
\hline\hline
2MASS ID & SpT & Assoc. & Ref$^{\rm a}$ & $J$ & $\delta$RA & $\delta$Dec & $d$ & Method$^{\rm b}$ & Mult$^{\rm c}$ \\
 &  &  &  & (mag) & (mas/yr) & (mas/yr) & (pc) &  &  \\
\hline
\endfirsthead
\caption{continued.}\\
\hline\hline
2MASS ID & SpT & Assoc. & Ref$^{\rm a}$ & $J$ & $\delta$RA & $\delta$Dec & $d$ & Method$^{\rm b}$ & Mult$^{\rm c}$ \\
 &  &  &  & (mag) & (mas/yr) & (mas/yr) & (pc) &  &  \\
\hline
\endhead
\hline
\endfoot
\hline
\multicolumn{10}{l}{\scriptsize{$^a$Reference for the assocation allocation. BII: Re-evaluation in this work with BANYANII, based on ambiguous}}\\
\multicolumn{10}{l}{\scriptsize{classification in previous works. GBII: Re-evaluation based on GAIA data. G15: \citet{gagne2015}. K14: }}\\
\multicolumn{10}{l}{\scriptsize{\citet{kraus2014}. M13: \citet{malo2013}. M14: \citet{malo2014}.}}\\
\multicolumn{10}{l}{\scriptsize{$^b$Method of distance estimation. Plx: Parallax. Kin: Kinematics. Spec: Spectroscopic.}}\\
\multicolumn{10}{l}{\scriptsize{$^c$The number of components seen in the AstraLux images (i.e. 1 is single, etc).}}\\
\hline
\endlastfoot

J00125703-7952073	&	M3.4	&	THA	&	K14	&	9.7	&	80.9	&	-46.1	&	48	&	Kin	&	1	\\
J00144767-6003477	&	M3.5	&	THA	&	K14	&	9.7	&	91.3	&	-63.1	&	42	&	Kin	&	1	\\
J00152752-6414545	&	M1.5	&	THA	&	K14	&	9.3	&	80.2	&	-49.9	&	50	&	Kin	&	1	\\
J00172353-6645124	&	M2.5	&	bPMG	&	M14	&	8.56	&	104.3	&	-13.5	&	39	&	Plx	&	1	\\
J00235732-5531435	&	M4.0	&	THA	&	K14	&	11.1	&	91.9	&	-66.9	&	42	&	Kin	&	1	\\
J00273330-6157169	&	M3.5	&	THA	&	K14	&	10.3	&	87.5	&	-56.8	&	44	&	Kin	&	1	\\
J00275023-3233060	&	M3.5	&	bPMG	&	M14	&	8.88	&	97.8	&	-60.9	&	32.3	&	Plx	&	1	\\
J00302572-6236015	&	K7.9	&	THA	&	K14	&	8.4	&	95.5	&	-48.4	&	44	&	Kin	&	3	\\
J00332438-5116433	&	M2.4	&	THA	&	K14	&	9.9	&	94.7	&	-59.9	&	42	&	Kin	&	1	\\
J00340843+2523498	&	K7	&	ABMG	&	M14	&	8.48	&	83.4	&	-98.2	&	49	&	Kin	&	2	\\
J00393579-3816584	&	M1.8	&	THA	&	K14	&	8.8	&	100.2	&	-65.5	&	40	&	Kin	&	1	\\
J00421010-5444431	&	M3.0	&	THA	&	K14	&	9.8	&	89.4	&	-47.9	&	46	&	Kin	&	1	\\
J00485254-6526330	&	M3.2	&	THA	&	K14	&	10.4	&	82.3	&	-40.7	&	50	&	Kin	&	1	\\
J00514081-5913320	&	M4.1	&	THA	&	K14	&	11.3	&	98.0	&	-50.3	&	42	&	Kin	&	2	\\
J01024375-6235344	&	M3.8	&	THA	&	K14	&	9.6	&	89.0	&	-39.6	&	46	&	Kin	&	2	\\
J01033563-5515561	&	M5.1	&	THA	&	K14	&	10.2	&	100.3	&	-46.9	&	47.2	&	Plx	&	3	\\
J01134031-5939346	&	M4.0	&	THA	&	K14	&	10.0	&	96.0	&	-35.4	&	44	&	Kin	&	1	\\
J01211297-6117281	&	M4.1	&	THA	&	K14	&	11.3	&	80.7	&	-28.3	&	52	&	Kin	&	1	\\
J01220441-3337036	&	K7	&	THA	&	GBII	&	8.31	&	110.1	&	-57.3	&	38.5	&	Plx	&	1	\\
J01233280-4113110	&	M5.6	&	THA	&	K14	&	10.8	&	109.3	&	-54.8	&	38	&	Kin	&	1	\\
J01275875-6032243	&	M4.0	&	THA	&	K14	&	11.1	&	88.4	&	-30.8	&	48	&	Kin	&	1	\\
J01283025-4921094	&	M4.0	&	THA	&	K14	&	10.6	&	101.4	&	-41.7	&	42	&	Kin	&	1	\\
J01372322+2657119	&	K5	&	GBII	&	M14	&	8.43	&	118.0	&	-128.0	&	37.6	&	Plx	&	1	\\
J01372781-4558261	&	M5.4	&	THA	&	K14	&	11.1	&	116.0	&	-37.1	&	38	&	Kin	&	2	\\
J01375879-5645447	&	M3.8	&	THA	&	K14	&	10.4	&	92.5	&	-32.8	&	46	&	Kin	&	1	\\
J01484087-4830519	&	M1.5	&	ABMG	&	M14	&	9.19	&	111.2	&	-51.0	&	36	&	Kin	&	1	\\
J01521830-5950168	&	M2-3	&	THA	&	M14	&	8.94	&	107.8	&	-27.0	&	40	&	Kin	&	1	\\
J01540267-4040440	&	K7	&	COL	&	M14	&	9.78	&	50.3	&	-14.7	&	84	&	Kin	&	1	\\
J02045317-5346162	&	K5	&	THA	&	M14	&	10.44	&	95.6	&	-30.9	&	42	&	Kin	&	1	\\
J02125819-5851182	&	M3.5	&	THA	&	K14	&	9.3	&	88.4	&	-16.1	&	48	&	Kin	&	1	\\
J02153328-5627175	&	M4.9	&	THA	&	K14	&	11.9	&	98.6	&	-29.3	&	42	&	Kin	&	1	\\
J02180960-6657524	&	M4.5	&	THA	&	K14	&	10.8	&	99.0	&	-14.7	&	42	&	Kin	&	2	\\
J02192210-3925225	&	M5.9	&	THA	&	K14	&	11.4	&	111.8	&	-44.0	&	36	&	Kin	&	1	\\
J02303239-4342232	&	K5	&	COL	&	GBII	&	8.02	&	78.1	&	-7.6	&	48.5	&	Plx	&	2	\\
J02341866-5128462	&	M4.2	&	THA	&	K14	&	10.6	&	100.6	&	-17.5	&	42	&	Kin	&	1	\\
J02414683-5259523	&	K6	&	THA	&	GBII	&	7.58	&	97.8	&	-15.4	&	43.3	&	Plx	&	1	\\
J02420204-5359147	&	M4.3	&	THA	&	K14	&	10.8	&	97.0	&	-20.8	&	42	&	Kin	&	1	\\
J02423301-5739367	&	K5	&	THA	&	GBII	&	8.56	&	84.9	&	-9.2	&	48.5	&	Plx	&	1	\\
J02442137+1057411	&	M0	&	BPMG	&	BII	&	7.97	&	68.4	&	-37.4	&	34.9	&	Plx	&	2	\\
J02523096-1548357	&	M2.5	&	ABMG	&	M13	&	10.54	&	78.0	&	-92.0	&	42	&	Kin	&	1	\\
J02553178-5702522	&	M4.4	&	THA	&	K14	&	11.1	&	89.5	&	-5.8	&	46	&	Kin	&	1	\\
J02591904-5122341	&	M5.3	&	THA	&	K14	&	11.7	&	81.7	&	-14.7	&	48	&	Kin	&	1	\\
J03050556-5317182	&	M5.1	&	THA	&	K14	&	11.1	&	89.4	&	-11.3	&	44	&	Kin	&	1	\\
J03083950-3844363	&	M4.7	&	THA	&	K14	&	11.2	&	68.3	&	-11.0	&	58	&	Kin	&	1	\\
J03093877-3014352	&	M4.7	&	THA	&	K14	&	11.6	&	88.7	&	-35.9	&	44	&	Kin	&	1	\\
J03114544-4719501	&	M3.7	&	THA	&	K14	&	10.4	&	88.4	&	-4.0	&	44	&	Kin	&	1	\\
J03190864-3507002	&	K7	&	THA	&	GBII	&	8.58	&	88.7	&	-20.0	&	45.3	&	Plx	&	1	\\
J03241504-5901125	&	K7	&	COL	&	M14	&	9.55	&	37.8	&	10.5	&	90	&	Kin	&	2	\\
J03291649-3702502	&	M3.7	&	THA	&	K14	&	10.6	&	89.8	&	-20.8	&	42	&	Kin	&	1	\\
J03315564-4359135	&	K6	&	THA	&	GBII	&	8.3	&	88.7	&	-4.2	&	45.2	&	Plx	&	2	\\
J03320347-5139550	&	M2	&	COL	&	M13	&	10.23	&	37.1	&	10.8	&	88	&	Kin	&	2	\\
J03454058-7509121	&	M4	&	THA	&	M14	&	10.82	&	64.3	&	23.0	&	54	&	Kin	&	1	\\
J03494535-6730350	&	K7	&	COL	&	M14	&	9.85	&	41.6	&	19.5	&	81	&	Kin	&	1	\\
J03512287-5154582	&	M4.4	&	THA	&	K14	&	10.6	&	71.9	&	2.4	&	50	&	Kin	&	1	\\
J03561624-3915219	&	M4.5	&	THA	&	K14	&	10.5	&	68.6	&	-3.7	&	52	&	Kin	&	2	\\
J04013874-3127472	&	M4.7	&	THA	&	K14	&	12.0	&	59.3	&	-12.3	&	58	&	Kin	&	1	\\
J04074372-6825111	&	M2.6	&	THA	&	K14	&	10.4	&	57.8	&	22.0	&	60	&	Kin	&	1	\\
J04082685-7844471	&	M0	&	CAR	&	M14	&	9.28	&	55.7	&	42.8	&	53	&	Kin	&	1	\\
J04091413-4008019	&	M3.5	&	COL	&	M14	&	10.65	&	46.4	&	8.1	&	63	&	Kin	&	1	\\
J04133314-5231586	&	M1.7	&	THA	&	K14	&	10.0	&	65.7	&	14.8	&	50	&	Kin	&	1	\\
J04133609-4413325	&	M3.3	&	THA	&	K14	&	10.8	&	56.7	&	0.4	&	60	&	Kin	&	1	\\
J04274963-3327010	&	M4.5	&	THA	&	K14	&	11.2	&	61.8	&	-0.7	&	52	&	Kin	&	1	\\
J04363294-7851021	&	M4	&	ABMG	&	M14	&	10.98	&	32.8	&	47.4	&	56	&	Kin	&	1	\\
J04435860-3643188	&	M3.5	&	THA	&	K14	&	10.7	&	54.1	&	-2.1	&	55	&	Kin	&	1	\\
J04440099-6624036	&	M0.5	&	THA	&	M14	&	9.47	&	53.0	&	30.2	&	55	&	Kin	&	1	\\
J04470041-5134405	&	M2.4	&	THA	&	K14	&	10.1	&	54.9	&	14.4	&	55	&	Kin	&	1	\\
J04475779-5035200	&	M4.1	&	THA	&	K14	&	10.9	&	47.9	&	17.8	&	60	&	Kin	&	2	\\
J04480066-5041255	&	K7	&	THA	&	GBII	&	8.74	&	56.9	&	17.1	&	57.7	&	Plx	&	2	\\
J04514615-2400087	&	M3	&	ABMG	&	M13	&	10.56	&	41.8	&	-56.9	&	45	&	Kin	&	1	\\
J04515303-4647309	&	M0	&	COL	&	M14	&	9.8	&	29.0	&	14.5	&	76	&	Kin	&	1	\\
J04554034-1917553	&	M0.5	&	ABMG	&	M14	&	9.78	&	21.8	&	-66.1	&	49	&	Kin	&	1	\\
J05015881+0958587	&	M3	&	bPMG	&	M13	&	7.21	&	12.1	&	-74.4	&	24.9	&	Plx	&	2	\\
J05064991-2135091	&	M1+M3.5+M4	&	bPMG	&	GBII	&	7.05	&	46.6	&	-16.3	&	19.7	&	Plx	&	3	\\
J05111098-4903597	&	M3.5	&	COL	&	M14	&	10.64	&	33.0	&	20.4	&	62	&	Kin	&	2	\\
J05142736-1514514	&	M3.5	&	COL	&	M14	&	10.71	&	36.0	&	-16.0	&	63	&	Kin	&	2	\\
J05142878-1514546	&	M3.5	&	COL	&	M14	&	10.95	&	34.1	&	-14.2	&	58	&	Kin	&	1	\\
J05164586-5410168	&	M3	&	COL	&	M14	&	10.43	&	26.3	&	26.6	&	69	&	Kin	&	1	\\
J05224571-3917062	&	K7	&	FLD	&	GBII	&	8.31	&	0.3	&	6.0	&	213.2	&	Plx	&	1	\\
J05240991-4223054	&	M0.5	&	ABMG	&	M14	&	10.58	&	4.9	&	-13.3	&	52	&	Kin	&	2	\\
J05332558-5117131	&	K7	&	FLD	&	GBII	&	8.99	&	42.2	&	27.0	&	55.2	&	Plx	&	1	\\
J05395494-1307598	&	M3	&	COL	&	M14	&	10.6	&	26.4	&	-18.9	&	54	&	Kin	&	1	\\
J05432676-3025129	&	M0.5	&	COL	&	M14	&	10.41	&	11.6	&	-0.3	&	87	&	Kin	&	1	\\
J06012186-1937547	&	M3.5	&	COL	&	M13	&	11.37	&	10.4	&	-9.0	&	79	&	Kin	&	1	\\
J06135773-2723550	&	K5	&	FLD	&	GBII	&	9.74	&	1.2	&	4.8	&	369	&	Plx	&	1	\\
J06153953-8433115	&	M3	&	bPMG	&	BII	&	9.25	&	-32.2	&	107.0	&	31	&	Kin	&	2	\\
J06475229-2523304	&	K7	&	Amb	&	M13	&	8.35	&	22.4	&	-70.3	&	240	&	Spec	&	2	\\
J06511418-4037510	&	K5	&	FLD	&	GBII	&	8.17	&	-1.0	&	7.1	&	1492	&	Plx	&	1	\\
J07170438-6311123	&	M2	&	Amb	&	M13	&	9.73	&	-13.1	&	48.0	&	53	&	Spec	&	1	\\
J07343426-2401353	&	M3.5	&	ARG	&	M13	&	10.65	&	-22.0	&	30.0	&	71	&	Kin	&	2	\\
J07523324-6436308	&	K7	&	CAR	&	M14	&	9.7	&	-5.9	&	27.8	&	88	&	Kin	&	3	\\
J07540718-6320149	&	M3	&	CAR	&	M14	&	10.33	&	-10.8	&	30.0	&	80	&	Kin	&	2	\\
J08083927-3605017	&	K7	&	COL	&	M14	&	9.53	&	-7.2	&	8.4	&	90	&	Kin	&	2	\\
J08173829-6817162	&	M3.8	&	CAR	&	G15	&	10.36	&	-15.6	&	58.1	&	53	&	Kin	&	2	\\
J08173943-8243298	&	M4	&	bPMG	&	M14	&	7.47	&	-81.9	&	102.6	&	27	&	Kin	&	2	\\
J08422284-8345248	&	K7	&	CAR	&	M13	&	9.45	&	-49.5	&	91.2	&	41	&	Kin	&	1	\\
J08465879-7246588	&	K7	&	FLD	&	GBII	&	8.49	&	-71.5	&	56.3	&	45.4	&	Plx	&	1	\\
J09032434-6348330	&	M0.5	&	CAR	&	M14	&	9.57	&	-34.5	&	35.4	&	66	&	Kin	&	3	\\
J09331427-4848331	&	K7	&	FLD	&	GBII	&	8.94	&	-48.9	&	19.6	&	45.5	&	Plx	&	1	\\
J09353126-2802552	&	K7	&	Amb	&	M13	&	8.51	&	-49.4	&	-57.4	&	46	&	Spec	&	1	\\
J10120908-3124451	&	M4	&	TWA	&	M14	&	8.85	&	-74.8	&	-9.4	&	53.9	&	Plx	&	2	\\
J10182870-3150029	&	M0	&	Amb	&	GBII	&	8.87	&	-55.6	&	-19.5	&	63.9	&	Plx	&	1	\\
J10252092-4241539	&	M1	&	Amb	&	M14	&	9.5	&	-46.8	&	-2.2	&	58	&	Kin	&	1	\\
J10585054-2346206	&	M3.8	&	Amb	&	G15	&	10.3	&	-92.3	&	-15.5	&	42	&	Kin	&	1	\\
J11112820-2655027	&	M3.8	&	TWA	&	G15	&	10.33	&	-87.4	&	-29.7	&	43	&	Kin	&	1	\\
J11132622-4523427	&	M0	&	TWA	&	M14	&	9.41	&	-44.1	&	-8.1	&	96.2	&	Plx	&	1	\\
J11210549-3845163	&	M1	&	TWA	&	M13	&	9	&	-68.3	&	-12.1	&	52	&	Kin	&	1	\\
J11211723-3446454	&	M1	&	TWA	&	M14	&	8.43	&	-67.4	&	-17.0	&	55.6	&	Plx	&	2	\\
J11321831-3019518	&	M5	&	TWA	&	M14	&	9.64	&	-89.6	&	-25.8	&	46	&	Kin	&	1	\\
J11393382-3040002	&	M3.9	&	TWA	&	G15	&	9.98	&	-79.9	&	-30.0	&	48	&	Kin	&	1	\\
J11455177-5520456	&	K5	&	FLD	&	GBII	&	8.02	&	-99.6	&	-5.9	&	42.6	&	Plx	&	1	\\
J12072738-3247002	&	M3	&	TWA	&	M14	&	8.62	&	-72.7	&	-29.3	&	53.8	&	Plx	&	1	\\
J12092998-7505400	&	M3	&	ARG	&	M14	&	9.91	&	-65.3	&	-0.6	&	77	&	Kin	&	2	\\
J12103101-7507205	&	M4	&	Amb	&	M13	&	11.19	&	-66.5	&	-5.8	&	28	&	Spec	&	1	\\
J12153072-3948426	&	M1	&	TWA	&	GBII	&	8.17	&	-76.5	&	-26.7	&	51.8	&	Plx	&	1	\\
J12170465-5743558	&	K7	&	FLD	&	GBII	&	8.71	&	-90.3	&	-11.3	&	231	&	Plx	&	1	\\
J12242443-5339088	&	M5	&	Amb	&	M13	&	10.51	&	-180.0	&	-60.0	&	16	&	Spec	&	1	\\
J12313807-4558593	&	M3	&	TWA	&	M14	&	9.33	&	-64.4	&	-28.6	&	77.5	&	Plx	&	1	\\
J13213722-4421518	&	M0.5	&	Amb	&	M13	&	9.74	&	-34.9	&	-18.8	&	64	&	Spec	&	1	\\
J13412668-4341522	&	M3.5	&	Amb	&	M14	&	10.75	&	-107.0	&	-60.8	&	42	&	Kin	&	1	\\
J13545390-7121476	&	M2.5	&	bPMG	&	M14	&	8.55	&	-165.0	&	-132.7	&	21	&	Kin	&	1	\\
J13591045-1950034	&	M4.5	&	ARG	&	M13	&	8.33	&	-552.7	&	-183.1	&	10.7	&	Plx	&	1	\\
J14142141-1521215	&	K5	&	bPMG	&	M13	&	7.43	&	-199.9	&	-172.8	&	30.2	&	Plx	&	3	\\
J14252913-4113323	&	M2.5	&	Amb	&	M14	&	8.55	&	-46.8	&	-49.2	&	66.9	&	Plx	&	1	\\
J14284804-7430205	&	M1	&	ARG	&	M13	&	9.26	&	-62.2	&	-36.3	&	76	&	Kin	&	1	\\
J14563812-6623419	&	M1.5	&	ARG	&	M13	&	10.4	&	-60.9	&	-40.4	&	73	&	Kin	&	1	\\
J15163224-5855237	&	K7	&	ARG	&	M13	&	9.1	&	-42.8	&	-44.2	&	77	&	Kin	&	2	\\
J15244849-4929473	&	M2	&	ABMG	&	M14	&	8.16	&	-121.1	&	-238.9	&	23	&	Kin	&	1	\\
J16074132-1103073	&	M4	&	ABMG	&	M14	&	9.82	&	-64.0	&	-148.0	&	36	&	Kin	&	2	\\
J16572029-5343316	&	M3	&	bPMG	&	GBII	&	8.69	&	-21.3	&	-90.9	&	51.5	&	Plx	&	3	\\
J17080882-6936186	&	M3.5	&	THA	&	M14	&	9.06	&	-55.6	&	-80.2	&	49	&	Kin	&	2	\\
J17115853-2530585	&	M1	&	ARG	&	M13	&	9.9	&	-15.1	&	-34.0	&	65	&	Kin	&	1	\\
J17130733-8552105	&	M0	&	THA	&	M14	&	8.59	&	-35.2	&	-58.1	&	62	&	Kin	&	2	\\
J17150219-3333398	&	M0	&	bPMG	&	M14	&	7.92	&	7.6	&	-176.9	&	23	&	Kin	&	1	\\
J17165072-3007104	&	M2.5	&	FLD	&	BII	&	10.37	&	-8.0	&	-36.0	&	65	&	Spec	&	2	\\
J17243644-3152484	&	K7	&	FLD	&	BII	&	9	&	-7.0	&	-34.6	&	56	&	Spec	&	3	\\
J17275761-4016243	&	M4	&	Amb	&	M13	&	10.04	&	-13.1	&	-50.4	&	43	&	Spec	&	1	\\
J17292067-5014529	&	M3	&	bPMG	&	M14	&	8.87	&	-6.3	&	-63.5	&	64	&	Kin	&	2	\\
J17300060-1840132	&	M3.5	&	Amb	&	M13	&	9.92	&	-8.1	&	-39.3	&	45	&	Spec	&	1	\\
J17580616-2222238	&	M1	&	Amb	&	M13	&	9.72	&	-5.9	&	-44.1	&	61	&	Spec	&	1	\\
J18141047-3247344	&	K6	&	Amb	&	M14	&	8.07	&	3.3	&	-52.0	&	73	&	Kin	&	1	\\
J18142207-3246100	&	M1.5	&	bPMG	&	M14	&	9.44	&	7.3	&	-39.9	&	90	&	Kin	&	1	\\
J18151564-4927472	&	M3	&	bPMG	&	M14	&	8.92	&	8.3	&	-71.5	&	61	&	Kin	&	1	\\
J18202275-1011131	&	K5+K7	&	bPMG	&	M14	&	7.64	&	10.5	&	-33.9	&	61	&	Kin	&	2	\\
J18420694-5554254	&	M3.5	&	bPMG	&	M14	&	9.49	&	11.2	&	-81.4	&	54	&	Kin	&	1	\\
J18450097-1409053	&	M5	&	ARG	&	M14	&	8.47	&	46.0	&	-84.0	&	16	&	Kin	&	2	\\
J18465255-6210366	&	M1	&	bPMG	&	M14	&	8.75	&	14.6	&	-81.1	&	54	&	Kin	&	1	\\
J18504448-3147472	&	K7	&	bPMG	&	GBII	&	8.31	&	13.7	&	-75.7	&	49.7	&	Plx	&	1	\\
J18553176-1622495	&	M0.5	&	ABMG	&	M13	&	9.13	&	32.6	&	-175.6	&	34	&	Kin	&	1	\\
J18580415-2953045	&	M0	&	FLD	&	GBII	&	8.86	&	8.4	&	-54.1	&	78.3	&	Plx	&	1	\\
J19102820-2319486	&	M4	&	bPMG	&	M14	&	9.1	&	17.6	&	-51.6	&	67	&	Kin	&	2	\\
J19225071-6310581	&	M3	&	THA	&	M14	&	9.45	&	-10.7	&	-77.4	&	61	&	Kin	&	1	\\
J19233820-4606316	&	M0	&	bPMG	&	M14	&	9.11	&	17.9	&	-57.9	&	70	&	Kin	&	1	\\
J19243494-3442392	&	M4	&	bPMG	&	M14	&	9.67	&	23.2	&	-72.1	&	54	&	Kin	&	1	\\
J19312434-2134226	&	M2.5	&	ARG	&	M14	&	8.69	&	63.0	&	-110.1	&	26	&	Plx	&	1	\\
J19420065-2104051	&	M3.5	&	ABMG	&	M13	&	8.69	&	48.0	&	-276.0	&	21	&	Kin	&	1	\\
J19560294-3207186	&	M4	&	bPMG	&	M14	&	8.96	&	32.6	&	-61.0	&	55	&	Kin	&	2	\\
J19560438-3207376	&	M0	&	bPMG	&	GBII	&	8.71	&	32.6	&	-68.9	&	50.2	&	Plx	&	1	\\
J20013718-3313139	&	M1	&	bPMG	&	M14	&	9.15	&	27.1	&	-60.9	&	61	&	Kin	&	1	\\
J20072376-5147272	&	K6	&	ARG	&	GBII	&	8.16	&	87.7	&	-143.6	&	33.7	&	Plx	&	1	\\
J20220177-3653014	&	M4.5	&	ABMG	&	M13	&	10.71	&	72.0	&	-188.0	&	33	&	Kin	&	1	\\
J20223306-2927499	&	M3.5	&	ABMG	&	M13	&	10.41	&	32.0	&	-104.0	&	58	&	Kin	&	2	\\
J20330186-4903105	&	M5.2	&	bPMG	&	G15	&	10.11	&	115.1	&	-208.0	&	16.3	&	Kin	&	1	\\
J20333759-2556521	&	M4.5	&	bPMG	&	M14	&	9.71	&	51.8	&	-76.8	&	48.3	&	Plx	&	1	\\
J20395460+0620118	&	K7	&	FLD	&	GBII	&	7.92	&	93.0	&	-103.3	&	38	&	Plx	&	1	\\
J20405616-8245093	&	M3.6	&	ARG	&	G15	&	9.52	&	89.9	&	-124.3	&	33	&	Kin	&	1	\\
J20560274-1710538	&	K7+M0	&	bPMG	&	M14	&	7.85	&	59.3	&	-63.0	&	44	&	Kin	&	2	\\
J21100535-1919573	&	M2	&	bPMG	&	M14	&	8.11	&	88.6	&	-92.5	&	32	&	Kin	&	1	\\
J21103096-2710513	&	M5	&	bPMG	&	M13	&	11.2	&	60.3	&	-79.7	&	40	&	Kin	&	1	\\
J21130526-1729126	&	K6	&	ABMG	&	GBII	&	8.35	&	76.9	&	-147.6	&	37.2	&	Plx	&	1	\\
J21212873-6655063	&	K7	&	bPMG	&	GBII	&	7.88	&	116.3	&	-90.9	&	32.1	&	Plx	&	1	\\
J21334415-3453372	&	M1.5	&	Amb	&	M13	&	10.24	&	40.4	&	-72.5	&	69	&	Spec	&	1	\\
J21342935-1840372	&	M4	&	ARG	&	M13	&	10.05	&	79.0	&	-11.4	&	51	&	Kin	&	2	\\
J21370885-6036054	&	M3	&	THA	&	M14	&	9.64	&	41.3	&	-91.3	&	48	&	Kin	&	1	\\
J22440873-5413183	&	M4	&	THA	&	M14	&	9.36	&	70.9	&	-60.1	&	49	&	Kin	&	2	\\
J22470872-6920447	&	K6	&	FLD	&	GBII	&	8.89	&	66.2	&	-65.8	&	52	&	Plx	&	1	\\
J23124644-5049240	&	M4.4	&	THA	&	K14	&	9.1	&	77.6	&	-75.7	&	46	&	Kin	&	2	\\
J23204705-6723209	&	M5	&	THA	&	M14	&	9.99	&	80.0	&	-97.1	&	41	&	Kin	&	2	\\
J23301341-2023271	&	M3	&	FLD	&	GBII	&	7.2	&	313.9	&	-204.6	&	16.1	&	Plx	&	1	\\
J23332198-1240072	&	K5	&	ARG	&	M13	&	10.28	&	164.2	&	9.0	&	31	&	Kin	&	2	\\
J23424333-6224564	&	M4.4	&	THA	&	K14	&	11.3	&	80.9	&	-61.6	&	46	&	Kin	&	1	\\
J23524562-5229593	&	M5.3	&	THA	&	K14	&	11.6	&	76.4	&	-82.4	&	44	&	Kin	&	1	\\

\label{t:sample}
\end{longtable}
}


{\scriptsize
\begin{longtable}{llrrrll}
\caption{Astrometric properties.}\\
\hline\hline
2MASS ID & Pair & $\rho$ & $\theta$ & Epoch & CPM$^{\rm a}$ & Ref$^{\rm b}$ \\
  &  & ($^{\prime \prime}$) & (deg) & (yr) &  &  \\
\hline
\endfirsthead
\caption{continued.}\\
\hline\hline
2MASS ID & Pair & $\rho$ & $\theta$ & Epoch & CPM$^{\rm a}$ & Ref$^{\rm b}$ \\
  &  & ($^{\prime \prime}$) & (deg) & (yr) &  &  \\
\hline
\endhead
\hline
\endfoot
\hline
\multicolumn{7}{l}{\scriptsize{$^a$Flag for CPM. Y if proven CPM, N if non-CPM, U if undetermined, I if implied (see text).}}\\
\multicolumn{7}{l}{\scriptsize{$^b$CPM reference. TP: This paper. WDS: The Washington Double Star catalogue \citep{mason2001}. B15:}}\\
\multicolumn{7}{l}{\scriptsize{\citet{bowler2015}. C10: \citet{chauvin2010}. D99: \citep{delfosse1999}. D13: \citep{delorme2013}. E15: }}\\
\multicolumn{7}{l}{\scriptsize{\citet{elliott2015}. J01:\citet{jayawardhana2001}.}}\\
\multicolumn{7}{l}{\scriptsize{$^c$Pair displays false triple effect -- possibly subject to a 180 deg ambiguity.}}\\
\endlastfoot

J00302572-6236015	&	AB$^{\rm c}$	&	0.098$\pm$0.005	&	262.4$\pm$0.6	&	2015.99	&	I	&	TP	\\
J00302572-6236015	&	AC	&	4.413$\pm$0.023	&	238.8$\pm$0.2	&	2015.99	&	Y	&	TP	\\
J00340843+2523498	&	AB	&	1.523$\pm$0.009	&	102.5$\pm$0.2	&	2015.91	&	Y	&	WDS	\\
J00514081-5913320	&	AB	&	10.990$\pm$0.052	&	260.2$\pm$0.2	&	2015.99	&	N	&	TP	\\
J01024375-6235344	&	AB	&	0.671$\pm$0.003	&	325.5$\pm$0.2	&	2015.99	&	I	&	TP	\\
J01033563-5515561	&	AB$^{\rm c}$	&	0.192$\pm$0.001	&	56.9$\pm$0.4	&	2015.98	&	Y	&	D13	\\
J01033563-5515561	&	AB--C	&	1.767$\pm$0.014	&	335.9$\pm$0.5	&	2015.98	&	Y	&	D13	\\
J01372781-4558261	&	AB	&	0.147$\pm$0.003	&	299.7$\pm$1.1	&	2015.99	&	I	&	TP	\\
J02180960-6657524	&	AB$^{\rm c}$	&	0.306$\pm$0.002	&	346.1$\pm$0.2	&	2015.99	&	I	&	TP	\\
J02303239-4342232	&	AB	&	0.097$\pm$0.001	&	212.5$\pm$1.2	&	2015.98	&	Y	&	E15	\\
J02442137+1057411	&	AB	&	0.249$\pm$0.002	&	322.6$\pm$0.2	&	2015.91	&	I	&	TP	\\
J03241504-5901125	&	AB	&	0.477$\pm$0.008	&	280.7$\pm$0.3	&	2015.98	&	U	&	N/A	\\
J03315564-4359135	&	AB	&	0.400$\pm$0.006	&	94.6$\pm$0.3	&	2015.99	&	Y	&	E15	\\
J03320347-5139550	&	AB	&	3.171$\pm$0.018	&	115.7$\pm$0.2	&	2015.98	&	U	&	N/A	\\
J03561624-3915219	&	AB	&	0.274$\pm$0.003	&	107.9$\pm$0.2	&	2015.99	&	U	&	N/A	\\
J04475779-5035200	&	AB$^{\rm c}$	&	0.530$\pm$0.002	&	121.4$\pm$0.2	&	2015.91	&	U	&	N/A	\\
J04480066-5041255	&	AB	&	1.047$\pm$0.011	&	345.4$\pm$0.2	&	2015.98	&	Y	&	E15	\\
J05015881+0958587	&	AB	&	1.398$\pm$0.012	&	146.9$\pm$0.2	&	2015.98	&	Y	&	D99	\\
J05064991-2135091	&	AB	&	8.454$\pm$0.039	&	307.0$\pm$0.2	&	2015.99	&	Y	&	WDS	\\
J05064991-2135091	&	BC	&	0.966$\pm$0.008	&	112.8$\pm$0.2	&	2015.99	&	Y	&	WDS	\\
J05111098-4903597	&	AB	&	2.397$\pm$0.017	&	168.2$\pm$0.2	&	2015.99	&	U	&	N/A	\\
J05142736-1514514	&	AB	&	2.679$\pm$0.016	&	268.5$\pm$0.2	&	2015.98	&	U	&	N/A	\\
J05240991-4223054	&	AB	&	0.225$\pm$0.005	&	61.1$\pm$0.4	&	2015.98	&	U	&	N/A	\\
J06153953-8433115	&	AB	&	0.965$\pm$0.008	&	275.5$\pm$0.2	&	2015.98	&	I	&	TP	\\
J06153953-8433115	&	AB	&	0.974$\pm$0.008	&	275.4$\pm$0.2	&	2015.99	&	I	&	TP	\\
J06475229-2523304	&	AB	&	1.076$\pm$0.013	&	27.9$\pm$0.2	&	2015.99	&	Y	&	WDS	\\
J07343426-2401353	&	AB	&	0.579$\pm$0.003	&	127.1$\pm$0.2	&	2015.99	&	U	&	N/A	\\
J07343426-2401353	&	AB	&	0.580$\pm$0.003	&	127.1$\pm$0.2	&	2016.38	&	U	&	N/A	\\
J07523324-6436308	&	AB	&	1.425$\pm$0.012	&	211.0$\pm$0.2	&	2015.98	&	U	&	N/A	\\
J07523324-6436308	&	BC	&	0.151$\pm$0.002	&	330.1$\pm$0.7	&	2015.98	&	U	&	N/A	\\
J07523324-6436308	&	AB	&	1.440$\pm$0.012	&	210.6$\pm$0.2	&	2016.38	&	U	&	N/A	\\
J07523324-6436308	&	BC	&	0.162$\pm$0.001	&	329.9$\pm$0.7	&	2016.38	&	U	&	N/A	\\
J07540718-6320149	&	AB	&	0.854$\pm$0.011	&	156.8$\pm$0.2	&	2015.98	&	U	&	N/A	\\
J07540718-6320149	&	AB	&	0.859$\pm$0.004	&	156.6$\pm$0.2	&	2016.38	&	U	&	N/A	\\
J08083927-3605017	&	AB	&	0.108$\pm$0.001	&	309.4$\pm$0.8	&	2015.99	&	U	&	N/A	\\
J08083927-3605017	&	AB	&	0.104$\pm$0.001	&	305.5$\pm$0.4	&	2016.38	&	U	&	N/A	\\
J08173829-6817162	&	AB	&	1.585$\pm$0.012	&	39.4$\pm$0.2	&	2016.38	&	U	&	N/A	\\
J08173943-8243298	&	AB	&	0.584$\pm$0.003	&	3.8$\pm$0.2	&	2015.98	&	Y	&	C10	\\
J08173943-8243298	&	AB	&	0.584$\pm$0.003	&	3.6$\pm$0.2	&	2016.38	&	Y	&	C10	\\
J09032434-6348330	&	AC	&	7.986$\pm$0.038	&	153.3$\pm$0.2	&	2015.98	&	U	&	N/A	\\
J09032434-6348330	&	AB	&	1.113$\pm$0.011	&	66.0$\pm$0.2	&	2015.98	&	U	&	N/A	\\
J09032434-6348330	&	AB	&	1.139$\pm$0.012	&	67.3$\pm$0.2	&	2016.38	&	U	&	N/A	\\
J10120908-3124451	&	AB	&	1.040$\pm$0.012	&	87.2$\pm$0.2	&	2015.98	&	Y	&	WDS	\\
J10120908-3124451	&	AB	&	1.037$\pm$0.011	&	86.9$\pm$0.2	&	2016.38	&	Y	&	WDS	\\
J11211723-3446454	&	AB	&	5.049$\pm$0.025	&	327.3$\pm$0.2	&	2015.98	&	Y	&	WDS	\\
J12092998-7505400	&	AB$^{\rm c}$	&	0.300$\pm$0.002	&	169.1$\pm$0.4	&	2015.99	&	Y	&	TP	\\
J12092998-7505400	&	AB	&	0.297$\pm$0.002	&	169.5$\pm$0.2	&	2016.38	&	Y	&	TP	\\
J14142141-1521215	&	AB	&	0.244$\pm$0.002	&	164.0$\pm$0.7	&	2016.38	&	Y	&	C10	\\
J14142141-1521215	&	AC	&	1.565$\pm$0.012	&	78.8$\pm$0.2	&	2016.38	&	Y	&	C10	\\
J15163224-5855237	&	AB	&	2.331$\pm$0.014	&	27.0$\pm$0.2	&	2016.38	&	U	&	N/A	\\
J16074132-1103073	&	AB	&	0.740$\pm$0.011	&	149.2$\pm$0.2	&	2016.38	&	Y	&	B15	\\
J16572029-5343316	&	AB	&	3.196$\pm$0.013	&	101.5$\pm$0.2	&	2016.38	&	U	&	N/A	\\
J16572029-5343316	&	AC	&	3.487$\pm$0.014	&	298.4$\pm$0.2	&	2016.38	&	U	&	N/A	\\
J17080882-6936186	&	AB	&	0.441$\pm$0.002	&	9.3$\pm$0.2	&	2016.38	&	I	&	TP	\\
J17130733-8552105	&	AB$^{\rm c}$	&	0.360$\pm$0.003	&	114.5$\pm$0.2	&	2016.38	&	U	&	N/A	\\
J17165072-3007104	&	AB	&	1.701$\pm$0.012	&	298.7$\pm$0.2	&	2016.38	&	U	&	N/A	\\
J17243644-3152484	&	AB	&	0.142$\pm$0.001	&	351.8$\pm$1.8	&	2016.38	&	U	&	N/A	\\
J17243644-3152484	&	AC	&	0.902$\pm$0.011	&	198.2$\pm$0.2	&	2016.38	&	U	&	N/A	\\
J17292067-5014529	&	AB	&	0.686$\pm$0.010	&	15.7$\pm$0.2	&	2016.38	&	Y	&	E15	\\
J18202275-1011131	&	AB	&	0.998$\pm$0.011	&	17.7$\pm$0.2	&	2016.38	&	Y	&	WDS	\\
J18450097-1409053	&	AB	&	3.224$\pm$0.016	&	123.8$\pm$0.2	&	2016.38	&	Y	&	TP	\\
J19102820-2319486	&	AB	&	0.424$\pm$0.002	&	172.8$\pm$0.2	&	2016.38	&	U	&	N/A	\\
J19560294-3207186	&	AB	&	0.177$\pm$0.001	&	250.7$\pm$0.2	&	2016.38	&	Y	&	B15	\\
J20223306-2927499	&	AB$^{\rm c}$	&	0.156$\pm$0.001	&	214.2$\pm$2.0	&	2016.38	&	I	&	TP	\\
J20560274-1710538	&	AB	&	2.181$\pm$0.013	&	140.3$\pm$0.2	&	2016.38	&	Y	&	J01	\\
J21342935-1840372	&	AB$^{\rm c}$	&	0.334$\pm$0.006	&	270.0$\pm$0.4	&	2016.38	&	I	&	TP	\\
J22440873-5413183	&	AB	&	0.517$\pm$0.003	&	298.3$\pm$0.2	&	2015.99	&	Y	&	C10	\\
J23124644-5049240	&	AB	&	0.426$\pm$0.003	&	71.4$\pm$0.2	&	2015.99	&	I	&	TP	\\
J23204705-6723209	&	AB	&	2.870$\pm$0.017	&	15.2$\pm$0.2	&	2015.99	&	Y	&	TP	\\
J23332198-1240072	&	AB	&	3.767$\pm$0.022	&	14.6$\pm$0.2	&	2015.99	&	N	&	TP	\\

\label{t:astro}
\end{longtable}
}


{\scriptsize
\begin{longtable}{llrrrrr}
\caption{Photometric and derived properties.}\\
\hline\hline
2MASS ID & Pair & $\Delta z^{\prime}$ & $m_1$ & $m_2$ & $a$ & $P_{\rm est}$ \\
  &   & (mag) & ($M_{\rm sun}$) & ($M_{\rm sun}$) & (AU) & (yr) \\
\hline
\endfirsthead
\caption{continued.}\\
\hline\hline
2MASS ID & Pair & $\Delta z^{\prime}$ & $m_1$ & $m_2$ & $a$ & $P_{\rm est}$ \\
  &   & (mag) & ($M_{\rm sun}$) & ($M_{\rm sun}$) & (AU) & (yr) \\
\hline
\endhead
\hline
\endfoot
\hline

\hline
\endlastfoot

J00302572-6236015	&	AC	&	1.33$\pm$0.02	&	0.6	&	0.6	&	194	&	2470	\\
J00340843+2523498	&	AB	&	0.04$\pm$0.01	&	0.6	&	0.6	&	75	&	589	\\
J00514081-5913320	&	AB	&	2.35$\pm$0.04	&	0.13	&	0.03	&	462	&	24792	\\
J01024375-6235344	&	AB	&	0.56$\pm$0.02	&	0.4	&	0.3	&	31	&	205	\\
J01372781-4558261	&	AB	&	1.30$\pm$0.05	&	0.11	&	0.06	&	6	&	32	\\
J02180960-6657524	&	AB	&	0.76$\pm$0.02	&	0.15	&	0.1	&	13	&	92	\\
J02442137+1057411	&	AB	&	2.03$\pm$0.05	&	0.7	&	0.2	&	9	&	27	\\
J03241504-5901125	&	AB	&	2.97$\pm$0.13	&	0.8	&	0.2	&	43	&	281	\\
J03315564-4359135	&	AB	&	4.09$\pm$0.27	&	0.7	&	0.08	&	18	&	87	\\
J03320347-5139550	&	AB	&	1.57$\pm$0.02	&	0.6	&	0.3	&	279	&	4914	\\
J03561624-3915219	&	AB	&	1.00$\pm$0.01	&	0.3	&	0.15	&	14	&	80	\\
J04480066-5041255	&	AB	&	0.92$\pm$0.02	&	0.7	&	0.5	&	60	&	429	\\
J05015881+0958587	&	AB	&	0.79$\pm$0.02	&	0.6	&	0.4	&	35	&	205	\\
J05064991-2135091	&	AB	&	0.49$\pm$0.01	&	0.6	&	0.5	&	162	&	1972	\\
J05064991-2135091	&	BC	&	0.53$\pm$0.01	&	0.5	&	0.4	&	19	&	84	\\
J05111098-4903597	&	AB	&	6.10$\pm$0.07	&	0.3	&	0.015	&	149	&	3228	\\
J05142736-1514514	&	AB	&	1.26$\pm$0.02	&	0.3	&	0.13	&	169	&	3344	\\
J05240991-4223054	&	AB	&	1.33$\pm$0.03	&	0.4	&	0.2	&	12	&	52	\\
J06153953-8433115	&	AB	&	0.13$\pm$0.01	&	0.17	&	0.15	&	30	&	289	\\
J06475229-2523304	&	AB	&	3.22$\pm$0.05	&	1.4	&	0.7	&	258	&	2864	\\
J07343426-2401353	&	AB	&	0.37$\pm$0.01	&	0.3	&	0.2	&	41	&	373	\\
J07523324-6436308	&	AB	&	2.41$\pm$0.03	&	0.7	&	0.2	&	125	&	1480	\\
J07523324-6436308	&	BC	&	0.40$\pm$0.06	&	0.2	&	0.17	&	13	&	80	\\
J07540718-6320149	&	AB	&	0.28$\pm$0.02	&	0.5	&	0.4	&	68	&	595	\\
J08083927-3605017	&	AB	&	1.39$\pm$0.06	&	0.7	&	0.4	&	10	&	29	\\
J08173829-6817162	&	AB	&	0.53$\pm$0.03	&	0.3	&	0.2	&	84	&	1089	\\
J08173943-8243298	&	AB	&	0.83$\pm$0.02	&	0.5	&	0.3	&	16	&	70	\\
J09032434-6348330	&	AB	&	2.37$\pm$0.02	&	0.6	&	0.15	&	73	&	727	\\
J09032434-6348330	&	AC	&	1.64$\pm$0.03	&	0.6	&	0.2	&	527	&	13529	\\
J10120908-3124451	&	AB	&	0.01$\pm$0.04	&	0.3	&	0.3	&	56	&	542	\\
J11211723-3446454	&	AB	&	0.00$\pm$0.02	&	0.7	&	0.7	&	281	&	3975	\\
J14142141-1521215	&	AB	&	3.03$\pm$0.21	&	0.8	&	0.15	&	7	&	21	\\
J14142141-1521215	&	AC	&	1.32$\pm$0.03	&	0.8	&	0.5	&	47	&	285	\\
J15163224-5855237	&	AB	&	0.00$\pm$0.03	&	0.7	&	0.7	&	179	&	2032	\\
J16074132-1103073	&	AB	&	0.08$\pm$0.03	&	0.3	&	0.3	&	27	&	178	\\
J16572029-5343316	&	AB	&	5.60$\pm$0.36	&	0.7	&	0.03	&	165	&	2471	\\
J16572029-5343316	&	AC	&	6.59$\pm$0.69	&	0.7	&	0.02	&	180	&	2836	\\
J17080882-6936186	&	AB	&	1.09$\pm$0.05	&	0.6	&	0.3	&	22	&	106	\\
J17165072-3007104	&	AB	&	0.99$\pm$0.02	&	0.5	&	0.3	&	111	&	1300	\\
J17243644-3152484	&	AB	&	2.29$\pm$0.09	&	0.7	&	0.4	&	8	&	21	\\
J17243644-3152484	&	AC	&	0.17$\pm$0.02	&	0.7	&	0.7	&	51	&	303	\\
J17292067-5014529	&	AB	&	0.18$\pm$0.02	&	0.6	&	0.6	&	44	&	266	\\
J18202275-1011131	&	AB	&	0.35$\pm$0.02	&	0.9	&	0.8	&	61	&	364	\\
J18450097-1409053	&	AB	&	0.11$\pm$0.02	&	0.2	&	0.2	&	52	&	586	\\
J19102820-2319486	&	AB	&	0.83$\pm$0.03	&	0.7	&	0.5	&	28	&	138	\\
J19560294-3207186	&	AB	&	1.71$\pm$0.04	&	0.6	&	0.2	&	10	&	34	\\
J20560274-1710538	&	AB	&	2.08$\pm$0.02	&	0.8	&	0.3	&	96	&	896	\\
J22440873-5413183	&	AB	&	0.40$\pm$0.01	&	0.4	&	0.3	&	25	&	152	\\
J23124644-5049240	&	AB	&	0.54$\pm$0.02	&	0.5	&	0.4	&	20	&	91	\\
J23204705-6723209	&	AB	&	1.09$\pm$0.02	&	0.3	&	0.15	&	118	&	1903	\\
J23332198-1240072	&	AB	&	6.60$\pm$0.08	&	0.3	&	0.02	&	117	&	2231	\\

\label{t:phot}
\end{longtable}
}

\end{document}